\documentclass[debug,usenatbib]{rmaa}
\usepackage{paralist}
\usepackage{longtable}

\usepackage{psfrag,color}

\title{Connecting the formation of stars and planets.\\ I -- Spectroscopic characterization of host stars with TIGRE } 

\author{
  L. M. Flor-Torres, \altaffilmark{1} 
  R. Coziol,\altaffilmark{1}
  K.-P. Schr\"oder,\altaffilmark{1}
  D. Jack,\altaffilmark{1}
  J. H. M. M. Schmitt,\altaffilmark{2}
  and S. Blanco-Cuaresma\altaffilmark{3}}

\altaffiltext{1}{Departamento de Astronom\'{\i}a, Universidad de Guanajuato, Guanajuato, Gto, M\'exico.}
\altaffiltext{2}{Hamburger Sternwarte, Universit\"at Hamburg, Hamburg, Germany.}
\altaffiltext{3}{Harvard-Smithsonian Center for Astrophysics, Cambridge, MA, USA.}

\shortauthor{Flor-Torres et al.}
\shorttitle{RevMexAA Main Journal Demo Document}

\fulladdresses{
\item L. M. Flor-Torres, R. Coziol, K.-P. Schr\"oder, D. Jack : Departamento de Astronom\'{\i}a, Universidad de Guanajuato, Guanajuato, Gto, M\'exico.
\item J. H. M. M. Schmitt: Hamburger Sternwarte, Universit\"at Hamburg, Hamburg, Germany.
\item S. Blanco-Cuaresma: Harvard-Smithsonian Center for Astrophysics, Cambridge, MA, USA.}
\indexauthor{Flor-Torres, L. M.}
\indexauthor{Coziol, R.}
\indexauthor{Schr\"oder, K.-P.}
\indexauthor{Schmitt, J. H. M. M.}
\indexauthor{Blanco-Cuaresma, S.}
\abstract{
In search for a connection between the formation of stars and the formation of planets, a new semi-automatic spectral analysis method using \textsf{iSpec} was developed for the TIGRE telescope installed in Guanajuato, Mexico. TIGRE is a 1.2m  robotic telescope, equipped with an Echelle spectrograph (HEROS), with a resolution R $\simeq 20000$. \textsf{iSpec} is a synthetic spectral fitting program for stars that allows to determine in an homogeneous way their fundamental parameters: effective temperature, $T_{\rm eff}$, surface gravity, $\log g$, metallicities, $[M/H]$ and $[Fe/H]$, and rotational velocity, $V \sin i$. In this first article we test our method by analysing the spectra of 46 stars, host of exoplanets, obtained with the TIGRE.  
}

\resumen{
En la b\'usqueda de una conexi\'on entre la formaci\'on estelar y planetaria, se desarroll\'o un nuevo m\'etodo semiautom\'atico de an\'alisis espectral estelar usando \textsf{iSpec}, para el telescopio TIGRE, instalado en Guanajuato, M\'exico. El TIGRE es un telescopio rob\'otico de 1.2m, el cu\'al est\'a equipado con el espectr\'ografo Echelle HEROS, que tiene una resoluci\'on R $\sim 20,000 $. \textsf{iSpec} es un programa de ajuste espectral sint\'etico para estrellas que permite determinar de manera homog\'enea sus par\'ametros fundamentales: temperatura efectiva, $ T_{\rm eff} $, gravedad superficial, $\log g $, metalicidades, $[M/H]$ y $[Fe/H]$, y velocidad de rotaci\'on, $V \sin i $. En este art\'iculo, probamos nuestro m\'etodo, analizando una muestra de 46 estrellas observadas por el TIGRE que albergan exoplanetas.}

\addkeyword{stars: fundamental parameters}
\addkeyword{stars: rotation}
\addkeyword{stars: formation}
\addkeyword{stars: planetary systems}

\begin{document}
\maketitle

\section{Introduction}
\label{sec:intro}

Since the discovery of the first planet orbiting another star in the 1990s, the number of confirmed exoplanets had steadily increased reaching in November of last year 4133.\footnote{http://exoplanet.eu/} The urgent tasks with which we are faced now are determining the compositions of these exoplanets and understanding how they formed. However, although that should have been straightforward \citep{Seager2010}, the detection of new types of planets had complicated the matter, changing in a crucial way our understanding of the formation of planetary systems around stars like the Sun.  

The first new type of planets to be discovered was the ``hot Jupiters'' \citep[HJs;][]{Mayor1995}, which are gas giants like Jupiter and Saturn, but with extremely small periods, $P < 10$\ days, consistent with semi-major axes smaller than $a_p = 0.05$\ AU. The existence of HJs is problematic because according to the model of formation of the solar system they can only form in the protoplanetary disk (PPD) where it is cold enough for volatile compounds such as water, ammonia, methane, carbon dioxide and monoxide to condense into solid ice grains \citep{Plummer2005}. In the solar system, this happens beyond the ice-line, which is located close to 3 AU \citep{Martin2012}. This implies that HJs must have formed farther out in the cold regions of the PPD, then migrated close to their stars \citep{Lin1996}. Subsequent discoveries have then shown that far from being exceptional, artifacts of an observational bias, HJs turned out to be very common around Sun-like stars, suggesting that large scale migration is a standard feature of the planet formation process \citep{Butler2000,Udry2007}. 

Two other new types of planets discovered are the ``Super-Earths'' \citep{Leconte2009,Valencia2006,Martin2015,Chabrier2009} and the ``mini-Neptunes'' \citep{Gandolfi2017}. These too were found to be common and very close to their stars, which, consequently, also makes them ``hot''. Their discoveries are important for two reasons. The first reason is that it makes the alternative ``in situ''  model for the formation of HJs \citep[e.g.,][]{Boss1997} a special model, since it cannot explain the large mass range and diversity of the ``hot'' exoplanets observed \citep[Super-Earths and mini-Neptunes in situ models are discussed in][]{Raymond2008,Chiang2013}. The second reason is that it was recently established by \citet{Lee2017} that their numbers around their host stars fall rapidly for periods $P < 10$ days ($\sim 0.09$ AU), which, assuming Keplerian orbits, clearly implies they all formed farther out (beyond 0.1 AU) and have migrated inward, but with a good many disappearing into their stars. This, once again, put large scale migration at the front scene of the planet formation process.  

This brings us to the present fundamental question in planet formation theory \citep{Anand2004}: what explains the fact that large scale migration did not happen in the solar system? Or, in other words, assuming all planets form in a PPD around a low mass star \citep{Nomura2016,vanderMarel2018,Perez2019}, what difference would make migration more important in one case and less important in another \citep[see discussion in][]{Walsh2011}? 

Integrating the migration process into a consistent model of planets formation is an extremely active and fast evolving field of research \citep[a recent review of this important subject can be find in][]{Raymond2020}. In the cases of the HJs, two migration mechanisms are accepted now as most probable \citep{Dawson2018}:\begin{inparaenum}[(1)]
\item disk migration, where the planet forms beyond the ice-line then migrates inward by loosing its orbit angular momentum to the PPD \citep[see thorough reviews in][]{Baruteau2014, Armitage2020}, and 
\item high-eccentricity migration, according to which the planet first gains a high eccentricity through interactions with other planets, which makes it to pass very close to its star where it looses its orbit angular momentum by tidal interactions \citep[this is a more complicated process, involving different mechanisms; e.g.,][]{Rasio1996,Weidenschilling1996,Marzari2002,Chatterjee2008,Nagasawa2008,Beauge2012}. 
\end{inparaenum} 
However, what is not clear in these two models is what importance must be put on the characteristics of the PPD, its mass, size, depth and composition? 

According to PPD formation theory, there are two possible mass scenarios \citep{Armitage2020}: the minimum mass model, between 0.01 to 0.02 M$_\odot$, which suggests that the PPD initial mass is only sufficient to explain the masses of the planets that formed within it, and the maximum mass model, which suggests the mass could have been much higher, close to 0.5 M$_\odot$. Consequently, more massive PPD (compared to the Solar system) might have either favored the formation of more massive planets \citep[consistent with PPD observations, see Fig. 2 and discussion in][]{Raymond2020} or a higher number of planets. The problem is that this makes both migration mechanisms equally probable \citep[also, the masses observed seem too low; also related to Fig. 2 in][]{Raymond2020}. Another caveat is that the solar system is a multiple planet system where migration on large scale did not happen. 

In terms of angular momentum, the differences between the minimum and maximum mass model for the PPD might also be important. By definition, the angular momentum of a planet is given by the relation \citep[e.g.,][]{Berget2010}:
\begin{equation}\label{eq1}
J_p = {M_p} \sqrt{G M_{*} a_p (1-e^2_p)}
\end{equation}
where  $M_{p}$ and $M_{*}$ are the masses of the planet and its host star,  $a_p$ is the  semi-major axis of the planet and $e_p$ its eccentricity. This suggests that within the maximum mass model more massive planets would also be expected to have higher orbital angular momentum (through their PPD) and, consequently, to have lost a larger amount of their angular momentum during large scale migration ($a_p \rightarrow 0$). This implies that the efficiency of the migration mechanism must increase with the mass of the planet (or its PPD). In principle, such requirement might be one way to distinguish which migration process is more realistic. However, the problem is bounded to be more complicated. First, stars rotate much more slowly than expected assuming conservation of angular momentum during their formation \citep[][]{McKee2007}. Second, defining the angular momentum of a planetary system as $J_{sys} = J_{*} + \Sigma J_{p}$
where $\Sigma J_{p}$ is the sum of the angular momentum of all the planets and $J_{*}$ the angular momentum of the host star \citep[cf.][]{Berget2010}, the angular momentum of massive planets  (even after migration, assuming $a_p \neq 0$) will always dominate over the angular momentum of its host stars. That is, $J_{*}/\Sigma J_{p} < 1$, and this is despite the enormous lost of angular momentum of the star during its formation. This implies that a sort of coupling must exist between the angular momentum of the stars and their planets through their PPDs. Understanding the nature of this coupling, therefore, is an important step in understanding how the PPD and the planets forming in it are connected to the formation of their stars. This, on the other hand, requires completing our information about the stars and the planets rotating around them. 

In the case of the planets, the two most successful detection techniques, the radial velocity (RV) and transit (Tr) methods, yield estimates of the mass of a planet, $M_p$, and its radius, $R_p$, as well as the  semi-major axis, $a_p$, and the eccentricity of its orbit, $e_p$. The first two parameters constrain their composition and formation process in the PPD, while the last two give information about their migration. By combining the four parameters we can also retrieve the angular momentum of the orbits of the planets (cf. Eq.~\ref{eq1}). In the case of the stars the most important characteristics that can be derived from their spectra are the effective temperature, $T_{eff}$, the surface gravity, $\log g$, the metallicities, [M/H] or [Fe/H], and the rotational velocity, $V \sin i$. The first two can be used in combination with their magnitudes and distances (using GAIA parallaxes) to determine their radii and masses, which taken in combination with the rotational velocity yield the angular momentum (or spin) of the star, $J_{*}$: 
\begin{equation}\label{eq2}
J_{*} = \gamma_{*} M_{*} R_{*} V^{rot}_{*}
\end{equation}
where $M_{*}$, $R_{*}$ and $\gamma_{*}$ are the star mass, radius and moment of inertia \citep[which depends on the mass of the star; cf.][]{Irwin2015}, and $V^{rot}_{*} =   V \sin i/\sin i$ is the equatorial rotation velocity (where $i$ is the inclination angle of the rotation axis relative to our line of sight).

To understand how the formation of planets is connected with the formation of their host stars, we must, consequently, make an effort in determining in parallel with the discovery of the former the physical characteristics of the latter. Present data banks for exoplanets (e.g., Kepler and now TESS, with 51 confirmed discoveries, and future survey like PLATO)\footnote{https://tess.mit.edu; about PLATO see https://platomission.com/about/} require a lot of follow up observations and analysis for the host stars, which are usually done using large diameter telescopes equipped with high resolution spectrographs. However, for the brightest stars (TESS targets, for example, being 30-100 times brighter than KEPLER stars), the use of smaller diameter telescopes equipped with lower resolution spectrographs might be more efficient in acquiring the information. Moreover, although high resolution spectra is justified when one uses the standard spectral analysis method, which is based on modeling the equivalent width (EW) of spectral lines, this might not be necessary when one uses the synthetic spectral analysis \citep[e.g.,][]{Valenti2005}, which consists in fitting observed spectra to grids of synthetic spectra with well determined physical characteristics that can be produced at different spectral resolutions. Another problem in using large aperture telescopes for host stars follow up is that since these telescopes are in high demand (for faint objects), data are collected on short duration runs by different groups using (although the same analysis method) different techniques and codes, which introduce discrepancies between the results \citep{Hinkel2014,Hinkel2016,Blanco-Cuaresma2014,Jofre2017}. This suggests that a follow up using a dedicated telescope and applying only one method of analysis could produce more homegeneous data \citep[one effort to homogenize data is the Stars With ExoplanETs CATalog or SWEET-Cat for short;][]{Sousa2008a}. For these reasons we developed a new method based on stellar spectral analysis for data obtained with the TIGRE telescope (Telescopio Internacional de Guanajuato Rob\'{o}tico Espectrosc\'{o}pico) that is installed at our institution in Guanajuato. 

TIGRE is a 1.2 m fully robotic telescope located at the La Luz Observatory (in central Mexico) at an altitude of 2,400 m; a more detailed description can be found in \citet{Schmitt2014}. Its principal instrument is the fibre-fed echelle spectrograph HEROS (Heidelberg Extended Range Optical Spectrograph), which yields a spectral resolution $R \sim 20,000$, covering a spectral range from 3800 \AA\  to 8800 \AA. The queue observing mode and automatic reduction pipeline already implemented for this telescope allow to optimized the observation and reduction process, producing high homogeneous data rapidly and confidently. To optimize the analysis process, we have developed a semi-automatic method that allows us to derive efficiently the most important physical characteristics of the stars: $T_{eff}$,  $\log g$, [M/H], [Fe/H], and  $V \sin i$. This was done by applying the synthetic spectral fitting technique as offered by the code \textsf{iSpec} \citep{Blanco-Cuaresma2014}, which was shown to yield results that are comparable to results in the literature that were obtained through different methods and codes \citep{Blanco-Cuaresma2019}.

The goal of this first article is to explain our spectral analysis method based on iSpec and compare results obtained by TIGRE with data taken from the literature. In an accompanying paper (Flor-Torres et al., hereafter paper {\rm II}) we will present a preliminary study, based on our own observational results, about the coupling of the angular momentum of the exoplanets and their host stars.

\section{Sample of host stars with exoplanets observed with TIGRE}
\label{sample}

Our initial target list for a pilot project was built from the revised compendium of confirmed exoplanets in the Exoplanet Orbit Database (hereafter Exoplanets.org,\footnote{http: //exoplanets.org/} selecting all the stars with spectral type F, G or K, located on the main sequence (based on their luminosities and colors), and for which a confirmed planet with well determined mass, radius, and semi-major axis was reported. Note that we did not apply a restriction to single systems, since from the point of view of the angular momentum we verified that only the major planet of a system counts (like Jupiter in our solar system). To optimize our observation with TIGRE, we restricted further our target list by retaining only host stars that have a magnitude V $ \leq 10.5$, obtaining a much shorter list of 65 targets. 

Our observed sample consists of 46 stars, host of 59 exoplanets, which were observed by TIGRE in queue mode. In Table~\ref{tab:0} the stars observed are given a running number (col.~1) which is used to identified them in the different graphics. The magnitude in V of each star and its distance as calculated from Gaia parallaxes are given in col.~3 and 4 respectively. Also shown are the exposure time, in col.~5, and signal to noise (S/N) in col.~6, as measured in the red part of the spectrum. The last column gives the main references found in the literature  with data about the host stars and their planetary systems.  

The HEROS spectrograph on TIGRE is coupled to two ANDOR CCDs, cooled by thermocouple (Peltier cooling to -100~C): blue iKon-L camera DZ936N-BBB  and red iKon-L camera DZ936N-BV. This yields for each star two spectra, one in the blue, covering a spectral range from 3800 \AA\  to 5750 \AA, and one in the red, covering a spectral range from 5850 \AA\ to 8750 \AA. All the data were automatically reduced by the TIGRE/HEROS standard pipeline, which applies automatically all the necessary steps to extract Echelle spectra \citep{Hempelmann2016,Mittag2016}: bias subtraction, flat fielding, cosmic ray correction, order definition and extraction and wavelength calibration, which was carried out by means of Th-Ar lamp spectra taken at the beginning and end of each night. Finally, we applied a barycentric correction and as a final reduction step, corrected each spectrum for telluric lines using the code \textsf{Molecfit} developed by \citet{Smette2015}. After verification of the results of the reduction process, we decided to concentrate our spectral analysis only on the red part of the spectra where the S/N is higher. 

\begin{table*}
 \caption[]{\small{Stars observed with the TIGRE}}
 \label{tab:0}
 \begin{scriptsize}
\begin{tabular}{ccccccc} 
\hline
\textbf{Id. \#} 	& \textbf{Star} 	&	\textbf{Magnitude}	&	\textbf{Distance}	& 	\textbf{Exp. time} & \textbf{S/N}  & \textbf{Ref.}\\
 							&  						&	(V)							&	(pc)						&  	(min)						 &					&	(as found in exoplanets.org)\\
\hline
\hline
1		& *KELT-6  	    &	10.3		&	242.4	&	97.1		&	54		& \citet{Damasso2015}  	\\
2		& *HD 219134 &	5.6   	&	6.5		&	8.0   	&	139		& \citet{Motalebi2015}	\\
3		& *KEPLER-37  &	9.8  		&	64.0		&	93.2   	&	75		& \citet{Batalha2013}	\\
4		& HD 46375  	&	7.8		&	29.6		&	108.0	&	107		& \citet{Marcy2000}	\\
5		& HD 75289  	&	6.4		&	29.1		&	37.8		&	99		& \citet{Udry2000}	\\
6		& HD 88133  	&	8.0		&	73.8		&	116.0	&	94		& \citet{Fischer2005a}	\\
7		& HD 149143  &	7.9		&	73.4		&	108.0	&	93		& \citet{Fischer2006,daSilva2006}   \\
8		& HAT-P-30  	&	10.4		&	215.3	&	100.9	&	59		& \citet{Johnson2011} 	\\
9		& KELT-3  	    &	9.8		&	211.3	&	92.5		&	68		& \citet{Pepper2013}	\\
10	& KEPLER-21  	&	8.3		&	108.9	&	29.4		&	83		& \citet{Borucki2011} 	\\
11	& KELT-2A  	   	&	8.7		&	134.6	&	54.3		&	95		& \citet{Beatty2012} 	\\
12	& HD86081  	&	8.7		&	104.2	&	61.4		&	100		& \citet{Johnson2006}	\\
13	& WASP-74  	&	9.8		&	149.8	&	96.5		&	73		& \citet{Hellier2015} 	\\
14	& HD 149026  &	8.1		&	76.0		&	37.4		&	98		& \citet{Sato2005} 	\\
15	& HD 209458  &	7.6		&	48.4		&	40.0		&	98		& \citet{Henry2000,Charbonneau2000}   	\\
16	& BD-10 3166 &	10.0		&	84.6		&	100.8	&	72		& \citet{Butler2000}	\\
17	& HD 189733  &	7.6		&	19.8		&	33.1		&	102		& \citet{Bouchy2005}	\\
18	& HD 97658  	&	7.7		&	21.6		&	35.0		&	123		& \citet{Howard2011} 	\\
19	& HAT-P-7  	   	&	10.5		&	344.5	&	43.5		&	32		& \citet{Pal2008} 	\\
20	& KELT-7  	    &	8.5		&	137.2	&	47.2		&	93		& \citet{Bieryla2015} 	\\
21	& HAT-P-14     &	10.0		&	224.1	&	84.0		&	57		& \citet{Torres2010} 	\\
22	& WASP-14   	&	9.7		&	162.8	&	74.6		&	66		& \citet{Joshi2009} 	\\
23	& HAT-P-2  	    &	8.7		&	128.2	&	70.0		&	69		& \citet{Bakos2007} 	\\
24	& WASP-38  	&	9.4		&	136.8	&	75.8		&	82		& \citet{Barros2011} 	\\
25	& HD 118203  &	8.1		&	92.5		&	41.5		&	92		& \citet{daSilva2006}	\\
26	& HD 2638  	    &	9.4		&	55.0		&	104.6	&	82		& \citet{Moutou2005}	\\
27	& WASP-13  	&	10.4		&	229.0	&	123.7	&	51		& \citet{Skillen2009} 	\\
28	& WASP-34  	&	10.3		&	132.6	&	136.8	&	62		& \citet{Smalley2011} 	\\
29	& WASP-82  	&	10.1		&	277.8	&	98.1		&	51		& \citet{West2016}	\\
30	&	HD17156		& 8.2	    & 78.3		& 46.3    	& 98	       & \citet{Fischer2007} \\
31	& XO-3  			&	9.9		&	214.3   &	70.8		&	60  		& \citet{Johns-Krull 2008} 	\\
32	& HD 33283  	&	8.0		&	90.1		&	53.4		&	101		& \citet{Johnson2006}	\\
33	& HD 217014  &	5.5		&	15.5		&	40.0		&	254		& \citet{Mayor1995}	\\
34	& HD 115383  &	5.2	   	&	17.5		&	4.0	   	&	105		& \citet{Kuzuhara2013}	\\
35	& HAT-P-6  	   	&	10.5		&	277.5	&	125.0	&	49		& \citet{Noyes2008}	\\
36	& *HD 75732 	&	6.0	    &	12.6		&	28.7		&	141		& \citet{Marcy2002}	\\
37	& HD 120136 	&	4.5	    &	15.7		&	9.3	    &	174		& \citet{Butler2000}	\\
38	& WASP-76 		&	9.5	  	&	195.3	&	91.1		&	73		& \citet{West2016} 	\\
39	& Hn-Peg  	    &	6.0		&	18.1		&	8.0		&	99		&	\citet{Luhman2007}\\
40 	& WASP-8  		& 	9.9 		& 90.2 	& 150.0 	& 	81 	 	& \citet{Queloz2010}	\\
41 	& WASP-69  	& 	9.9  		& 50.0 	&  90.0  	& 	76 	 	& \citet{Anderson2014}	\\
42 	& HAT-P-34  	& 	10.4  	& 251.1 	& 105.0  	& 	56 	 	& \citet{Bakos2012}	\\
43 	& HAT-P-1  		&	 9.9  	& 159.7 	&  75.0  	& 	60 	 	& \citet{Bakos2007} 	\\
44 	& WASP-94 A  	& 	10.1  	& 212.5 	& 105.0  	& 	58 	 	& \citet{Neveu-VanMalle2014}	\\
45	& WASP-111  	& 	10.3  	& 300.5 	&  90.0  	& 	58 	 	& \citet{Anderson2014}	\\
46 	& HAT-P-8  		& 	10.4  	& 212.8 	& 150.0  	& 	74 	 	& \citet{Latham2009}	\\
\hline
\multicolumn{7}{l}{An * in front of the name of the star identified multiple planetary systems.}
\end{tabular}
 \end{scriptsize}
 \end{table*}
 
 \begin{figure}
\includegraphics[width=0.7\linewidth, angle=270]{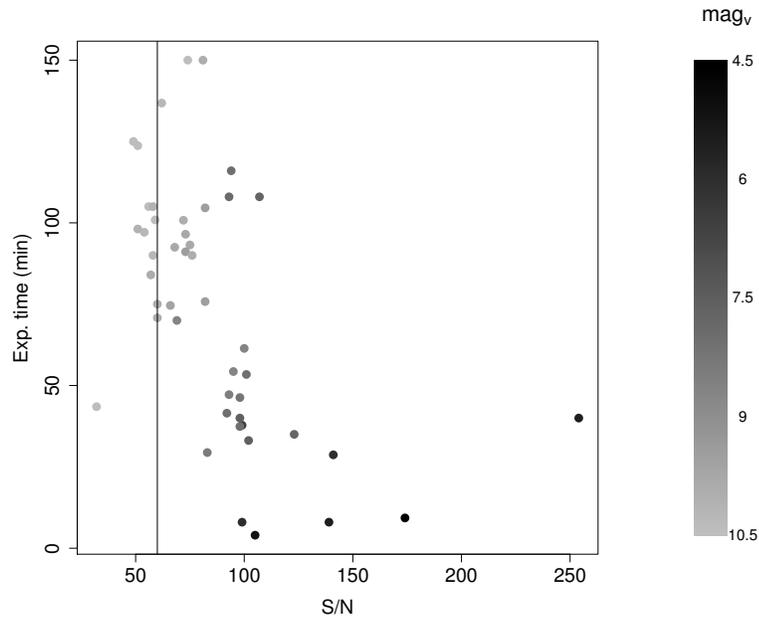}
\caption{\textit{S/N as a function of exposure time for our sample, limited to stars with magnitude limit V~$ \leq 10.5$. Note that the exposure time was adjusted to reach S/N~$\geq~60$ in less than two hours.}}
\label{sn_exptime}
\end{figure}

 \begin{figure}
\includegraphics[width=0.8\linewidth, angle=0]{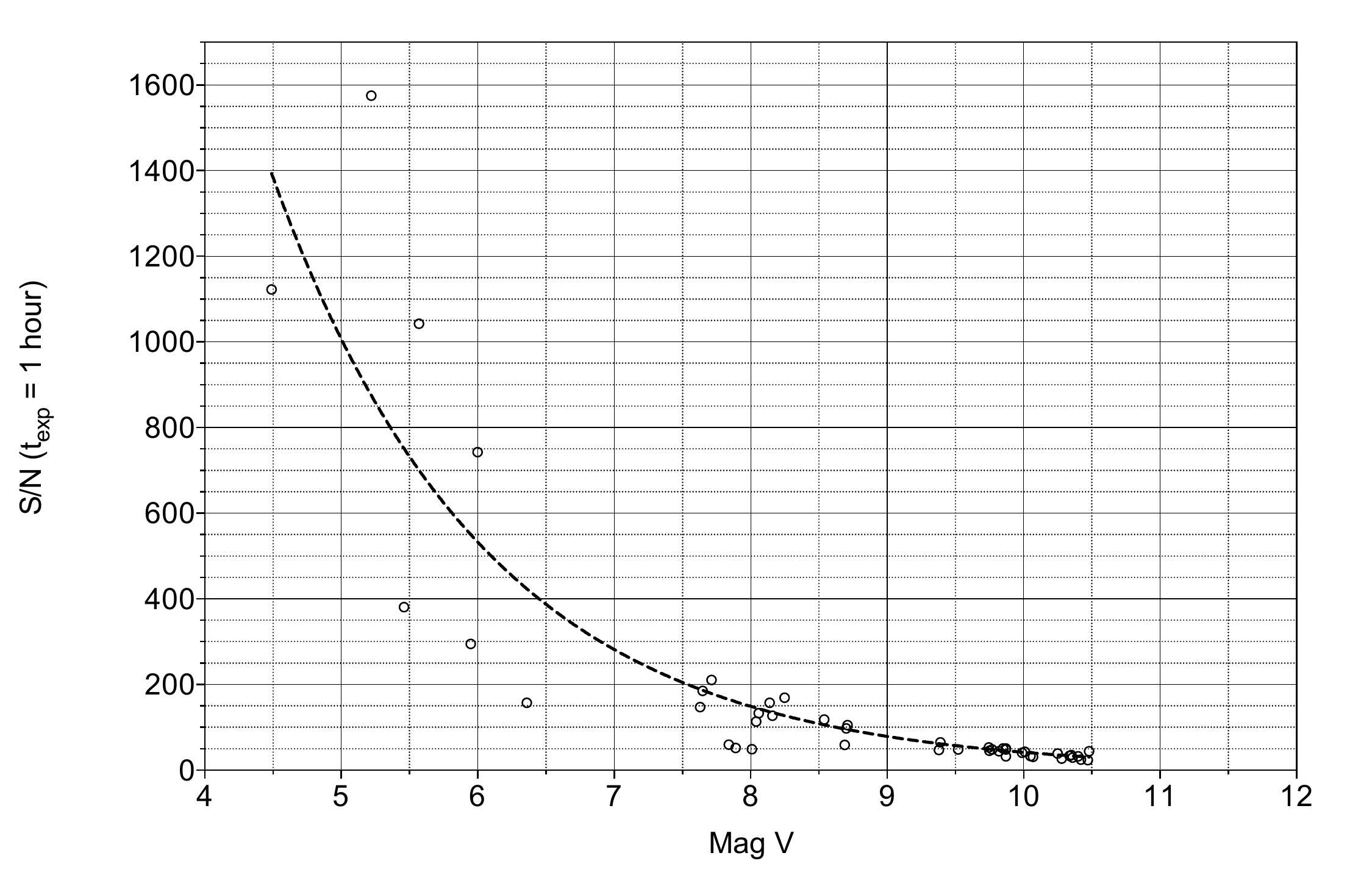}
\caption{\textit{Exponential growth curve giving the S/N expected after one hour exposure time for stars with different magnitudes.}}
\label{SNvsV}
\end{figure}
 
In Fig.~\ref{sn_exptime} we traced the S/N obtained as a function of the exposure time. For each star the total exposure time during observation was adjusted to reach S/N $\geq 60$. Note that this result only depends on the telescope diameter, the fiber transmission, the spectrograph resolution (we used R = 20,000, but the resolution is adjustable in \textsf{iSpec}) and the photometric conditions (explaining most of the variance).The average exposure time was 74~s for an average S/N~$\sim 87$, which makes observation with TIGRE a very efficient process. 

To determine how faint a follow up with TIGRE could be done efficiently, we traced in Fig.~\ref{SNvsV} an exponential growth curve based on our data, determining the S/N expected in one hour for stars with different magnitudes. One can see that a star with 10.5 mag in V would be expected to have a S/N near 30 (or 60 in 2 hours). The lower we could go would be S/N $\sim10$ which would be reached in one hour by a 12.5 mag star (or 2 hours for 13 mag star). Since it is not clear how low the S/N of a star could be to be efficiently analysed using the synthetic-spectra method, we judged safer to adopt a limit S/N of 60, which can be reach within two hours using TIGRE. This justify the magnitude limit, V~$\leq 10.5$,  adopted for this pilot project. Our observations suggest that a 1.2~m telescope could contribute significantly to follow up of exoplanet surveys like TESS, searching for small rocky planets around bright stars (stars much brighter than KEPLER stars), and in the near future PLATO, which will search for Earth-like planets in the habitable zones of one million nearby Solar type stars. 

\section{Spectral analysis using iSpec}
\label{method}

\begin{figure}
\includegraphics[width=0.7\linewidth, angle=270]{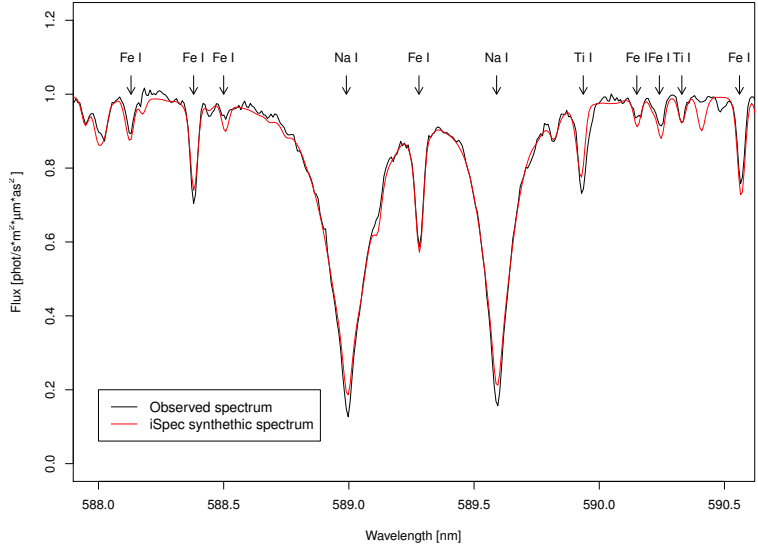}
\caption{\textit{Example of the result for the synthetic spectral fitting method in iSpec. The star is HD 46375, the observed spectrum is in blue and the fitted spectrum in red, with a rms of 0.0319.}}
\label{f1}
\end{figure}

Our spectral analysis was developed using the synthetic spectral fitting technique offered by the code \textsf{iSpec} \citep[version 2016.11.18;][]{Blanco-Cuaresma2014, Blanco-Cuaresma2019}. In brief, this technique consists in comparing an observed spectrum with synthetic spectra interpolated from pre-computed grids, calculated using different radiative transfer codes, and applying a least-square minimization algorithm to converge towards the closest approximation possible. In Fig.~\ref{f1} we show one example of a synthetic spectral fit for the star HD 46375. The fit has a rms~0.0319, which is relatively good considering HEROS intermediate resolution \citep{Piskunov2017}. Due to the low resolution of our spectra we can fit at the same time in an homogeneous manner the intensity and spectral profiles of more than 100 lines \citep[compared to a few 10s at high resolution; e.g.,][]{Valenti2005}. The best fit then allows to determine five important atmospheric parameters: i.e., the effective temperature, $T_{eff}$, the surface gravity, $\log g$, two indexes of metallicities, $[M/H]$ and  $[Fe/H]$, and the rotational velocity, $V \sin i$. 

To optimize our analysis a crucial step of our method consisted in applying \textsf{iSpec} to a TIGRE spectrum from the Sun (as reflected by the Moon). Our main goal was to determine a subset of spectral lines and segments that best reproduces the physical characteristics of our star. Although this step is time consuming, because each line and segment have to be tested incrementally by running \textsf{iSpec}, once these lines and segments are established, the analysis of stars becomes straightforward and efficient, the full process taking only a few minutes to converge on a modern desktop computer. Starting with the whole line-list available in the VALD database \citep{Kupka1999,Kupka2011}, we kept only 122 lines in the red for which we defined specific segments in Table~\ref{tab:B1} in the appendix~\ref{sec:ap-A}. As we already verified in \citet{Eisner2020}, these lines and segments can also be used in iSpec as a standard basis for observations obtained with different telescopes and (once adjusted for the resolution) other spectrographs.  

Our initial analysis of the Sun also allowed us to decide which solar abundance, atmospheric model and radiative transfer code were optimal. We adopted the solar abundance of \citet{Asplund2009}, the \texttt{ATLAS} atmospheric model of \citet{Kurucz2005} and the radiative transfer code \texttt{SPECTRUM} of \citet{Gray1994}. Another parameter that turned out to be important using \textsf{iSpec} is a correction for limb darkening, which we fixed to a value of 0.6  \citep{Hestroffer1998,Blanco-Cuaresma2019}. 

After working out the analysis of the Sun, we found an unexpected difficulty in obtaining the rotation velocity,  $V \sin i$, for our stars. The problem comes from the fact that in low mass stars the turbulence velocity $V_{mic}$ and $V_{mac}$ have values comparable to $V \sin i$ \citep{Doyle2014}, and there is consequently no fail-proof recipe how to ``constrain'' these velocities using the synthetic method. One way to approach this problem (following different researchers in the field) is in adopting ad hoc values based on theory or observation \citep{Gray1984a,Gray1984b,Fischer2005b,Bruntt2010,Tsantaki2014,Doyle2014}. For our analysis, we decided to adopt empirical values. For $V_{mac}$ we used the relation \citep{Doyle2014}:
\begin{equation}\label{eq4}
V_{mac} = a + b\Delta T + c\Delta T^2 -2.00 (\log g - 4.44)
\end{equation}
where $\Delta T = (T_{eff} - 5777)$, $ a = 3.21$, $ b = 2.33 \times 10^{-3}$ and $c = 2.00 \times 10^{-6}$. 
For $V_{mic}$ we used the relation \citep{Tsantaki2014}:
\begin{equation}\label{eq5}
V_{mic} = 6.932 \times 10^{-4} T_{eff} -0.348 \log g - 1.437 
\end{equation}  
Note that neither authors give uncertainties on these values. However, \citet{Doyle2014} suggested generic uncertainties of the order of $\pm\ 0.27$ km/s and $\pm\ 0.15$ for $V_{mac}$ and $V_{mic}$ respectively, which we adopted for our study. 

\begin{figure}
\includegraphics[width=0.7\linewidth, angle=270]{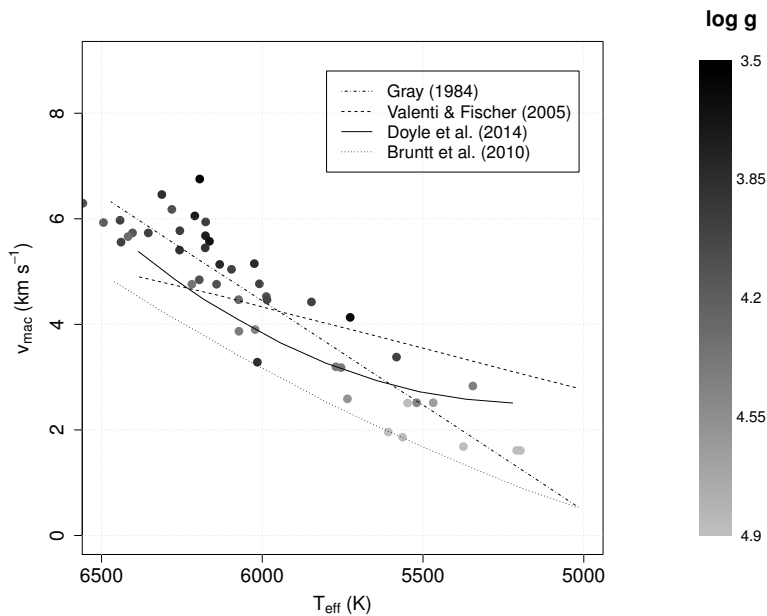}
\caption{\textit{Values of $V_{mac}$ adopted for our analysis with iSpec as function of our results for $T_{eff}$. }}
\label{Vmac}
\end{figure}

In Fig.~\ref{Vmac} we show the final values of $V_{mac}$ obtained in our analysis. Traced over the data, we draw the different relations proposed in the literature to fix this parameter. At high temperatures ($T_{eff} > 5800$ K), one can see that our values for $V_{mac}$ are well above the upper limit determined by \citep{Valenti2005}, while at low temperatures, the values are well above the lower limit determined by \citep{Bruntt2010}. In general, our results for $V_{mac}$ are consistent with the values expected based on the relation proposed by \citep{Gray1984b}.

Our final result for the Sun is shown in Table~\ref{tab:2}. These values were obtained after only ten iterations, using the parameters of the Sun as initial guess and fixing $V_{mac}$ and $V_{mic}$ using Eq.~\ref{eq4} and Eq.~\ref{eq5}. For comparison sake, we also included in Table~\ref{tab:2} the values adopted for the Gaia Benchmark stars. Although our best fit reproduces well the physical characteristics of the Sun, the uncertainty  estimated by \textsf{iSpec} for $V \sin i$ is relatively high. But this, as we already explained, is expected considering the problem related to $V_{mic}$ and $V_{mac}$. The different solutions (as shown in Fig.~\ref{Vmac}) in solving this problem might explain, for example, why the macro turbulence we used for the Sun is lower than what was used by Gaia. In \citet{Doyle2014}, the authors already noted a similar difference, by comparing the values they obtained by their relation with results reported by \citet{Fischer2005b}, where $V_{mac}$ were systematically higher by as much as $0.54\ {\rm km\ s^{-1}}$. However, adding this difference (as a systematic correction) to make our result for $V_{mac}$ closer to the value proposed in the Gaia Benchmark, did not lower the uncertainties on $V \sin i$ obtained with \textsf{iSpec}. Therefore, considering that our method easily reproduce the value of $V \sin i$ for the Sun, we judged more realistic to keep a high uncertainty on this parameter. Besides, the question is possibly more complex, considering the uncertainty on the existence of a $J-M$ relation,  $J_{*} \propto M^{\alpha}$, for low mass stars \citep{Herbst2007} and taking into account that $V \sin i$ might also depend on the age of the star \citep[that is, decreasing with the age;][]{Kraft1967,Wilson1963,Skumanich1972}. 

\begin{table}[h!]
 \centering
 \caption[]{\small{Results for the Solar spectrum using iSpec}}
\label{tab:2}
  \vspace{0.25cm} 
 \begin{small}
\begin{tabular}{ccc}
\hline
\textbf{Char.} 	& \textbf{iSpec}  & \textbf{Sun*} 		\\
 	                      &  	\\
\hline
\hline
 $T_{eff}$	 	& $5571\pm30$ K 			& $5571$ K\\
 $\log g$ 	 	& $4.44\pm0.04$ dex 	& $4.44$ dex\\
 $[M/H]$			& $0.00\pm0.03$ 			& $0$ 			\\
 $[Fe/H]$ 		& $0.00\pm0.03$ 			& $0$ 			\\
 $V \sin i$		& $1.60\pm1.45$ km/s 	& $1.60$ km/s   \\
 $V_{mic}$		& $1.02$ km/s 				& $1.07$ km/s   \\
 $V_{mac}$		& $3.19 $ km/s 				& $4.21$ km/s   \\
rms of fit 			&	$0.0289$ 					&  						\\
\hline
\multicolumn{3}{l}{*Gaia benchmark Stars values\citep{Blanco-Cuaresma2019}}
\end{tabular} 
\end{small}
 \end{table}

For the analysis of the stars, our semi-automatic method can be resumed in the following way. We first run \textsf{iSpec} using the parameters of the Sun as initial input. This implies calculating $V_{mac}$ and $V_{mic}$ using Eq.~\ref{eq4} and Eq.~\ref{eq5}, keeping these values fixed and leaving all the other parameters free. The results of the first run give us new values of $T_{eff}$ and $\log g$ based on which we calculate new initial values for $V_{mac}$ and $V_{mic}$ before running iSpec a second time. 

To verify our solutions, for each star we use the final value of $T_{eff}$ to calculate its mass and radius (first we get the mass, then the corresponding radius) using the mass-luminosity relation for stars with masses between 0.43 $M_\odot$ and 2 $M_\odot$ \citep{Wang2018}:
\begin{equation}\label{eq6}
\frac{M}{M_\odot} = (\frac{L}{L_\odot})^{1/4} = \frac{T_{eff}}{T_\odot} (\frac{R}{R_\odot})^{1/2}
\end{equation}
where $L$ is the bolometric luminosity as determined from its magnitude in V and distance calculated from Gaia in Table~\ref{tab:0}. Then, we verified that the value of $\log g$ given by \textsf{iSpec} is consistent with the mass and radius obtained using the relation \citep[Eq.~7 in][]{Valenti2005}: 
\begin{equation}\label{eq7}
\log(M/M_\odot) = \log(g_{*}) + 2\log(R/R_\odot) - 4.437
\end{equation}

In general, we obtained consistent values for $\log g$, within the generic errors suggested by \citet{Doyle2014}. However, for height stars, we found discrepant masses, the masses obtained using Eq.~\ref{eq7} being higher than the masses using Eq.~\ref{eq6}. To solve this problem we found important to  better constrain the initial value of $\log g$ before running \textsf{iSpec} a second time. The reason for this constrain is physically clear, since, as shown by Eq.~\ref{eq6} and Eq.~\ref{eq7}, $\log g$ is coupled to $T_{eff}$. In \citet{Valenti2005}, for example, the authors took into account this coupling by first fixing the initial value of $T_{eff}$ related to the B-V color of the star, then used a generic $\log g$ consistent with this temperature. In our case we decided to use as initial parameters for the second run the value of $T_{eff}$ obtained from the first run with iSpec (which use the values of the Sun as first guesses) and to use as second guess the value of $\log g$ given by Eq.~\ref{eq7} that makes the two masses consistent. This also implies recalculating $V_{mac}$ and $V_{mic}$ for this new values, which, as before, are kept fix running \textsf{iSpec}. The unique consequence of adding this constrain for the height stars with discrepant masses was to lower the final values of their $\log(g)$, all the other parameters being equal. For each star, our method requires only two runs of ten iterations each, which amounts to about 30 minutes CPU time on a fast desktop computer. This makes our analysis process quite efficient.  

\section{Results: characterization of the host stars of exoplanets observed with TIGRE}
\label{results}

\begin{table*}
 \begin{minipage}{1.3\textwidth}
 \centering
 \caption[]{\small{Physical parameters of the host stars of exoplanets in our sample, as determined with \textsf{iSpec}.}}
 \label{tab:3b}
\begin{Huge}
\resizebox{\textwidth}{!}{%
 \begin{tabular}{clccccccccccccccccc}
\hline
\textbf{No.} & Name of star & $T_{eff}$ & $\Delta{\rm T_{eff}}$ & $log g$ & $\Delta{\log g}$ & {[M/H]} & $\Delta{[M/H]}$ & $[Fe/H]$ & $\Delta{[Fe/H]}$ & $V \sin i $ & $\Delta{V \sin i}$ & $V_{mic}$ & $V_{mac}$ & rms & R$_{*}$ & $\Delta{\rm R_{*}}$ & M$_{*}$  & $\Delta{\rm M_{*}}$ \\ 
&& (K) & (K)&  &  & &&&&(km/s)&(km/s)&(km/s)&(km/s)& & ($R_\odot$) & ($R_\odot$) &  ($M_\odot$)&  ($M_\odot$) \\
\hline
1			& KELT-6 			& 6176	& 24	& 4.03	& 0.05	& $-0.38$	& 0.02	& $-0.14$	& 0.03& 6.52	& 0.82	& 1.44	& 5.28	& 0.0292	& 1.71	&	0.20	& 1.22	&	0.20	\\ 
2			& HD 219134	& 5209	& 13	& 4.90	& 0.00	& 0.00			& 0.02	& 0.02			& 0.01& 7.09	& 0.30	& 0.47	& 1.61	& 0.0318	& 0.54	&	0.09	& 0.77	&	0.09	\\ 
3			& KEPLER-37 	& 5520	& 19	& 4.50	& 0.04	& $-0.40$	& 0.02	& $-0.28$	& 0.02& 6.62	& 0.50	& 0.82	& 2.62	& 0.0317	& 0.71	&	0.15	& 0.88	&	0.15	\\ 
4			& HD 46375 	& 5345	& 22	& 4.47	& 0.04	& $-0.05$	& 0.01	& 0.11			& 0.01& 2.01	& 0.73	& 0.71	& 2.52	& 0.0319	& 0.83	&	0.01	& 0.88	&	0.02	\\ 
5			& HD 75289 	& 6196	& 23	& 4.16	& 0.06	& 0.16			& 0.02	& 0.42			& 0.02& 4.11	& 0.56	& 1.41	& 5.10	& 0.0291	& 1.27	&	0.01	& 1.14	&	0.03	\\
6			& HD 88133 	& 5582	& 16	& 4.05	& 0.03	& 0.15			& 0.01	& 0.34			& 0.01& 1.98	& 0.76	& 1.02	& 3.61	& 0.0344	& 1.80	&	0.01	& 1.12	&	0.02	\\ 
7			& HD 149143	& 6067	& 20	& 4.36	& 0.04	& 0.17			& 0.02	& 0.48			& 0.02& 3.53	& 0.61	& 1.25	& 4.22	& 0.0316	& 1.64	&	0.10	& 1.19	&	0.10	\\ 
8			& HAT-P-30 		& 6177	& 30	& 3.81	& 0.08	& 0.06			& 0.04	& 0.15			& 0.03& 8.88	& 0.60	& 1.52	& 5.72	& 0.0324	& 1.51	&	0.19	& 1.19	&	0.19	\\ 
9			& KELT-3 			& 6404	& 26	& 4.20	& 0.05	& 0.02			& 0.03	& 0.24			& 0.02& 8.51	& 0.57	& 1.54	& 5.93	& 0.0294	& 1.77	&	0.16	& 1.28	&	0.16	 \\
10		& KEPLER-21 	& 6256	& 31	& 4.02	& 0.06	& $-0.07$	& 0.03	& 0.11			& 0.03& 7.38	& 0.57	& 1.50	& 5.63	& 0.0317	& 1.96	&	0.10	& 1.28	&	0.10	\\ 
11		& KELT-2A 		& 6164	& 22	& 3.74	& 0.05	& $-0.04$	& 0.03	& 0.19			& 0.02& 7.28	& 0.51	& 1.53	& 5.81	& 0.0315	& 2.01	&	0.10	& 1.27	&	0.10	\\ 
12		& HD86081 		& 6015	& 19	& 3.94	& 0.06	& 0.14			& 0.02	& 0.38			& 0.02& 4.01	& 0.57	& 1.36	& 4.88	& 0.0314	& 1.63	&	0.13	& 1.18	&	0.13	 \\
13		& WASP-74 		& 5727	& 14	& 3.70	& 0.03	& 0.14			& 0.02	& 0.22			& 0.02& 8.24	& 0.38	& 1.25	& 4.58	& 0.0341	& 1.57	&	0.15	& 1.11	&	0.16	\\ 
14		& HD 149026 	& 6096	& 14	& 4.06	& 0.02	& 0.25			& 0.02	& 0.38			& 0.02& 5.28	& 0.49	& 1.38	& 4.92	& 0.0302	& 1.51	&	0.10	& 1.17	&	0.10	\\ 
15		& HD 209458 	& 5988	& 17	& 4.17	& 0.06	& $-0.22$	& 0.03	& $-0.01$	& 0.02& 2.96	& 0.86	& 1.26	& 4.33	& 0.0282	& 1.25	&	0.10	& 1.10	&	0.10	\\ 
16		& BD-10 3166	& 5578	& 23	& 4.64	& 0.04	& 0.22			& 0.01	& 0.39			& 0.02& 6.88	& 0.38	& 0.82	& 2.43	& 0.0361	& 0.82	&	0.17	& 0.92	&	0.17	 \\
17		& HD 189733 	& 5374	& 18	& 4.89	& 0.04	& $-0.04$	& 0.01	& 0.09			& 0.01& 2.75	& 0.60	& 0.59	& 1.70	& 0.0287	& 0.60	&	0.01	& 0.82	&	0.02	\\ 
18		& HD 97658 	& 5468	& 20	& 4.68	& 0.04	& $-0.39$	& 0.01	& $-0.17$	& 0.01& 1.87	& 0.85	& 0.72	& 2.20	& 0.0320	& 0.62	&	0.01	& 0.84	&	0.02	\\ 
19		& HAT-P-7 		& 6270	& 46	& 3.95	& 0.12	& $-0.01$	& 0.07	& 0.43			& 0.05& 5.70	& 1.44	& 1.53	& 5.82	& 0.0312	& 2.21	&	0.19	& 1.32	&	0.20	\\ 
20		& KELT-7 			& 6508	& 38	& 3.95	& 0.13	& $-0.19$	& 0.01	& 0.15			& 0.04& 45.2	& 1.39	& 1.70	& 6.96	& 0.0269	& 2.01	&	0.15	& 1.34	&	0.17	\\ 
21		& HAT-P-14 		& 6490	& 35	& 4.12	& 0.07	& $-0.11$	& 0.04	& 0.09			& 0.03& 8.86	& 0.65	& 1.63	& 6.53	& 0.0293	& 1.69	&	0.15	& 1.28	&	0.16	 \\
22		& WASP-14 		& 6195	& 24	& 3.60	& 0.04	& $-0.33$	& 0.03	&$ -0.23$	& 0.03& 1.47	& 2.11	& 1.60	& 6.21	& 0.0298	& 1.50	&	0.15	& 1.19	&	0.15	\\ 
23		& HAT-P-2 		& 6439	& 24	& 4.05	& 0.05	& 0.15			& 0.03	& 0.29			& 0.03& 20.66& 0.58	& 1.62	& 6.41	& 0.0254	& 1.79	&	0.10	& 1.29	&	0.10	\\ 
24		& WASP-38 		& 6178	& 18	& 3.95	& 0.04	& $-010$		& 0.03	& 0.15			& 0.02& 7.47	& 0.54	& 1.47	& 5.44	& 0.0301	& 1.49	&	0.13	& 1.18	&	0.13	 \\
25		& HD 118203 	& 5847	& 31	& 4.06	& 0.06	& 0.04			& 0.01	& 0.19			& 0.02& 4.16	& 0.58	& 1.20	& 4.14	& 0.0321	& 2.04	&	0.10	& 1.21	&	0.10	 \\
26		& HD 2638 		& 5564	& 18	& 4.90	& 0.00	& 0.14			& 0.02	& 0.38			& 0.02& 3.30	& 0.62	& 0.71	& 1.88	& 0.0355	& 0.72	&	0.13	& 0.89	&	0.13	 \\
27		& WASP-13 		& 6025	& 29	& 3.89	& 0.03	& $-0.01$	& 0.03	& 0.12			& 0.03& 2.35	& 1.30	& 1.39	& 5.01	& 0.0344	& 1.62	&	0.20	& 1.18	&	0.20	 \\
28		& WASP-34 		& 5771	& 27	& 4.44	& 0.04	& $-0.31$	& 0.03	& 0.00			& 0.03& 1.60	& 1.39	& 1.02	& 3.20	& 0.0326	& 1.08	&	0.20	& 1.02	&	0.20	 \\
29		& WASP-82 		& 6257	& 28	& 3.96	& 0.08	& $-0.05$	& 0.04	& 0.22			& 0.03& 2.86	& 1.23	& 1.52	& 5.75	& 0.0331	& 2.16	&	0.17	& 1.31	&	0.18	\\ 
30		& HD17156 		& 5985	& 22	& 4.10	& 0.05	& $-0.06$	& 0.01	& 0.09			& 0.02& 3.02	& 0.78	& 1.29	& 4.46	& 0.0303	& 1.58	&	0.10	& 1.16	&	0.10	\\ 
31		& XO-3 			& 6281	& 30	& 4.16	& 0.10	& $-0.12$	& 0.04	& $-0.19$	& 0.03& 20.2	& 0.73	& 1.47	& 5.45	& 0.0270	& 1.83	&	0.16	& 1.27	&	0.16	\\ 
32		& HD 33283 	& 5877	& 16	& 3.81	& 0.03	& 0.05			& 0.02	& 0.32			& 0.02& 4.39	& 0.47	& 1.31	& 4.72	& 0.0320	& 1.99	&	0.09	& 1.21	&	0.09	\\ 
33		& HD 217014 	& 5755	& 12	& 4.43	& 0.03	& $-0.30$	& 0.01	& $-0.01$	& 0.02& 0.40	& 1.43	& 1.01	& 3.18	& 0.0312	& 1.16	&	0.19	& 1.04	&	0.19	\\ 
34		& HD 115383 	& 5891	& 19	& 4.19	& 0.04	& $-0.16$	& 0.02	& 0.22			& 0.02& 8.11	& 0.40	& 1.19	& 4.00	& 0.0285	& 1.41	&	0.01	& 1.11	&	0.02	 \\
35		& HAT-P-6 		& 6442	& 34	& 4.05	& 0.05	& 0.04			& 0.02	& $-0.10$	& 0.03& 11.7	& 0.71	& 1.62	& 6.43	& 0.0440	& 1.70	&	0.17	& 1.28	&	0.18	 \\
36		& HD 75732 	& 5548	& 17	& 4.89	& 0.03	& 0.19			& 0.01	& 0.38			& 0.01& 0.17	& 1.61	& 0.71	& 1.88	& 0.0338	& 0.80	&	0.19	& 0.91	&	0.19	 \\
37		& HD 120136 	& 6210	& 17	& 3.79	& 0.04	& 0.20			& 0.02	& 0.20			& 0.02& 15.14& 0.36	& 1.55	& 5.89	& 0.0292	& 1.61	&	0.19	& 1.21	&	0.19	 \\
38		& WASP-76 		& 6133	& 21	& 3.90	& 0.04	& 0.10			& 0.02	& 0.40			& 0.02& 2.24	& 1.00	& 1.46	& 5.37	& 0.0301	& 2.03	&	0.16	& 1.27	&	0.16	 \\
39		& HN-PEG 		& 5853	& 18	& 4.41	& 0.04	& $-0.37$	& 0.02	& 0.03			& 0.02& 10.02& 0.41	& 1.09	& 3.46	& 0.0337	& 1.03	&	0.01	& 1.02	&	0.02	 \\
40		& WASP-8     	& 5735	& 55	& 4.62	& 0.13	& 0.10			& 0.02	& 0.39			& 0.04& 6.45	& 1.07	& 0.93	& 2.76	& 0.0308	& 0.89	&	0.19	& 0.97	&	0.20	\\ 
41		& WASP-69   	& 5197	& 15	& 4.90	& 0.00	& 0.22			& 0.01	& 0.30			& 0.01& 1.18	& 1.07	& 0.46	& 1.61	& 0.0360	& 0.58	&	0.15	& 0.78	&	0.15	\\ 
42		& HAT-P-34   	& 6494	& 33	& 4.22	& 0.07	& 0.14			& 0.04	& 0.38			& 0.04& 25.32& 0.80	& 1.60	& 6.35	& 0.0287	& 1.57	&	0.19	& 1.26	&	0.19	\\ 
43		& HAT-P-1   		& 6142	& 24	& 4.15	& 0.06	& 0.14			& 0.02	& 0.21			& 0.03& 5.65	& 0.66	& 1.38	& 4.91	& 0.0330	& 1.41	&	0.10	& 1.16	&	0.10	\\ 
44		& WASP-94A   	& 5988	& 23	& 3.76	& 0.04	& 0.17			& 0.03	& 0.38			& 0.02& 5.55	& 0.60	& 1.41	& 5.15	& 0.0338	& 1.80	&	0.17	& 1.20	&	0.18	\\ 
45		& WASP-111   	& 6312	& 32	& 3.94	& 0.08	& 0.05			& 0.04	& 0.30			& 0.03& 11.57& 0.54	& 1.57	& 6.03	& 0.0308	& 2.12	&	0.18	& 1.32	&	0.18	\\ 
46		& HAT-P-8    	& 6009	& 60	& 4.06	& 0.09	& 0.15			& 0.05	& $-0.12$	& 0.07& 13.68& 1.09	& 1.32	& 4.62	& 0.0365	& 1.55	&	0.16	& 1.16	&	0.16	\\
\hline
 \end{tabular}
}
\end{Huge}
 \end{minipage}
 \end{table*}
 
Our measurements for the physical parameters of the host stars as determined with our semi-automatic method are presented in Table~\ref{tab:3b}. Note that for the metallicities, $[M/H]$ and $[Fe/H]$, an extra correction was needed following  \citet[][]{Valenti2005}, to eliminate spurious abundance trends (see their explanations in section 6.4). This correction is based on the assumption that the ratio of one elemental abundance to another must not vary systematically with the temperature. The correction then is simple: it consists in tracing the metallicities as a function of $T_{eff}$, fitting a second order relation, then subtract this spurious relation from the data. All the uncertainties reported in Table~\ref{tab:3b} were calculated by \textsf{iSpec}, while the errors of the radii and masses are the quadratic sums of the uncertainties of the parameters used to calculate these values \citep[see subsection 7.2 in][]{Valenti2005}. As explained in \citet{Valenti2005} and in \citet{Piskunov2017} the uncertainties estimated by the algorithms that produce the synthetic spectra and fit them to the observed spectra are usually undetermined, as compared to the random errors calculated from the measurements of multiple observations of the stars \citep[subsection 6.3 in][]{Valenti2005}. Unfortunately, multiple observations were not programmed for our stars and we have only 4 stars in our study (17, 19, 23 and 46) that were observed more than once (four times for three and six times for the fourth one). This means that only a rough estimate of the random error can be obtained for our pilot-survey by calculating for each of these stars the standard deviations of the parameters measured applying the same spectral analysis as for the other stars. In table~\ref{tab:3a} we compare our mean uncertainties as obtained with \textsf{iSpec} with the mean of the standard deviations for the multiply observed stars in our sample. Except for $V \sin i$, the mean empirical errors are much higher than the \textsf{iSpec} values. In particular, our empirical errors are higher than the empirical uncertainties calculated by \citet[Figure 9 in][]{Valenti2005}, being comparable to their 2 sigma probabilities (the values in the table corresponds to 1 sigma, the threshold that includes 68.3\% of their error measurements).  

Comparing with Exoplanets.org and SWEET.cat mean uncertainties, our mean errors (standard deviations of multiple stars) are slightly higher, although still comparable to those reported in these studies. Although preliminary, this result is important as it suggests that our  results based on \textsf{iSpec} analysis of low resolution spectra (R~$\sim 20,000$) are in good agreement with results obtained using higher resolution spectra (R higher than 50,000). 

\begin{table}[h!]
 \centering
 \caption[]{\small{Comparison of errors}}
\label{tab:3a}
  \vspace{0.25cm} 
 \begin{small}
\begin{tabular}{lcccc}
\hline
\textbf{Errors} 	& \textbf{$T_{eff}$} & \textbf{$\log g$}	& \textbf{$[Fe/H]$}	& \textbf{$V \sin i$} 		\\
 	                      &  	(K) 						& (dex) 					& (dex) 						& (km/s) \\
\hline
\hline
Standard deviations			&	73 	&	0.14 	& 0.08 	& 0.8\\
\textsf{iSpec} 					&	25 	&	0.05 	& 0.02 	& 0.8 \\
Valenti \& Fischer 			& 44		&	0.06		&	0.03 	& 0.5\\
Exoplanets.org	       			&	66 	&	0.06 	& 0.07 	& 0.7\\
SWEET-Cat			   			&	52	&	0.10 	& 0.04 	& \\			
\hline
\end{tabular} 
\end{small}
 \end{table} 

 \begin{figure}[h!]
\includegraphics[width=1.15\linewidth, angle=0]{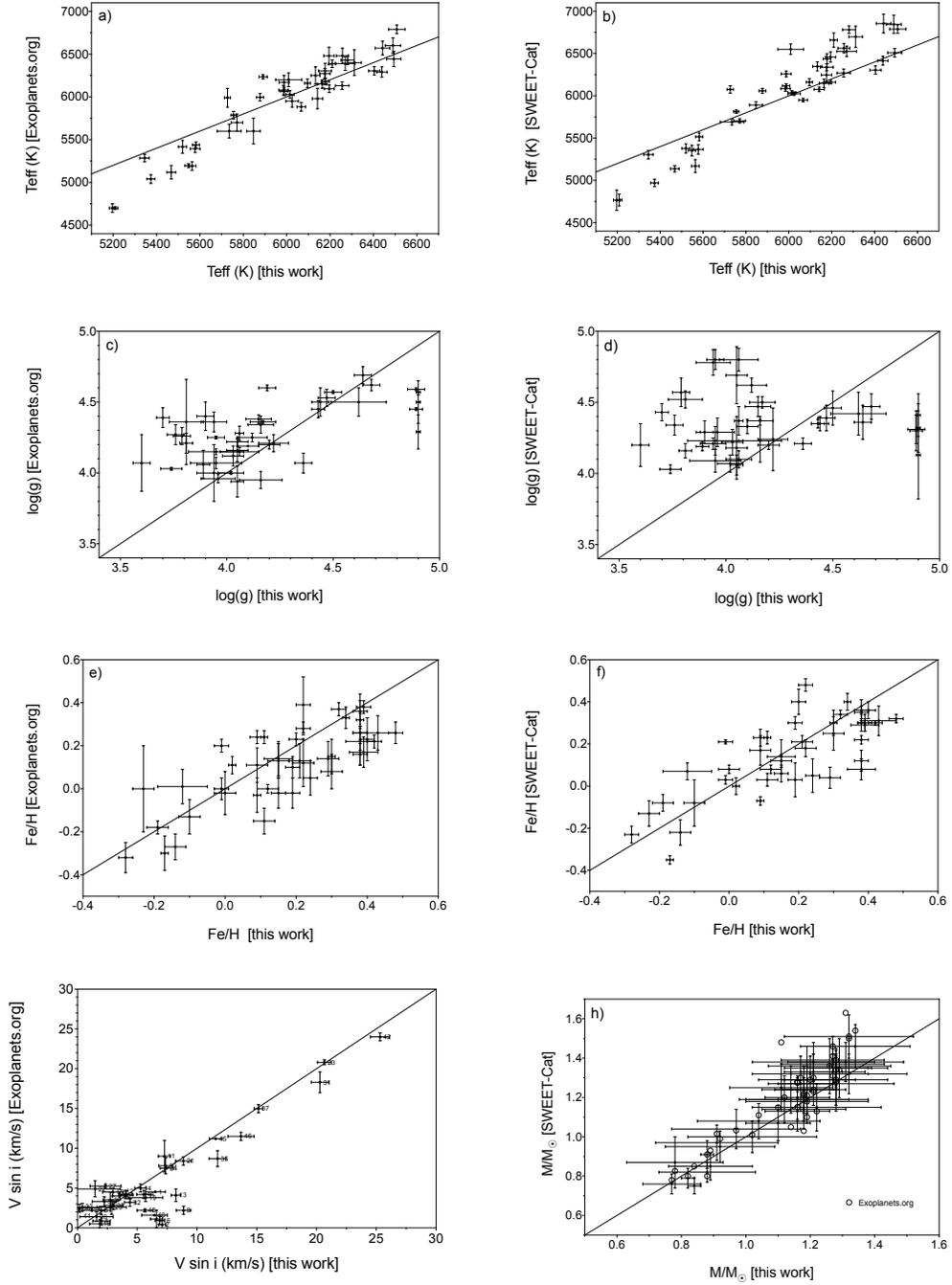}
\caption{\textit{Comparison of our results with those in Exoplanets.org (left) and SWEET-Cat (right): a) and b)  $T_{eff}$,  c) and d)  $\log g$, e) and f) $[Fe/H]$ g) $V \sin i$ h) the mass of the stars, $M_{*}$, with data for Exoplanets.org included.}}
\label{Comp}
\end{figure}

Another way to verify the consistency of our data is to compare our results with those published in Exoplanets.org (on the left in Fig.~\ref{Comp}) and SWEET-Cat (on the right). Taken as a whole, our results seem compatible with the data reported in these two catalogs (note that the uncertainties are those of \textsf{iSpec}), although there are also slight notable differences. In Fig.~\ref{Comp}a, our values for $T_{eff}$ are slightly higher below 5800 K than the values reported by Exoplanets.org and SWEET-Cat. However, above 6000 K our temperatures are comparable with those in Exoplanets.org, while clearly lower compared to SWEET-Cat. The largest difference between our results and those of the two other surveys is for $\log g$. Compared with Exoplanets.org (Fig.~\ref{Comp}c), our values for $\log g$ are comparable within the range 4-4.7 dex, only slightly underestimated. Above 4.7 dex, our values tend to be overestimated while below 4 dex they are underestimated. These differences are amplified comparing with SWEET-Cat in Fig.~\ref{Comp}d. Once again, however, we must conclude that these differences already existed comparing Exoplanets.org with SWEET-Cat. Despite the above differences, our metallicities in Fig.~\ref{Comp}e and Fig.~\ref{Comp}f are comparable with those published both by Exoplanets.org and SWEET-Cat. Once against our results seem more similar to the former than the latter. 

The most important comparison for the purpose of our survey is for $V \sin i$ in Fig.~\ref{Comp}g. Unfortunately, we can only compare with Exoplanets.org, since SWEET-Cat did not publish their results. What we find is a very good agreement, with only a slight trend for our values to be higher. This trend is most probably due to our lower resolution and the different way we determined $V_{mic}$ and $V_{mac}$ (more about that will be said later). In Fig.~\ref{Comp}h we compare the masses of the stars as compare to those reported by SWEET-Cat. This time we observe a much better consistency. Note that we have also included the values given by Exoplanets.org (as open circles). In general, our masses show a weak trend to be lower, although well within the uncertainties.

To quantify the differences between our values and those reported in Exoplanets.org and SWEET-Cat we compare in Table~\ref{tab:5} the medians and means (note that since the numbers of stars in the comparisons vary the means and medians are not the same). In both cases, we also determined if the differences are statistically significant, using non-parametric Mann-Whitney tests \citep{Dalgaard2008}. The last two columns in Table~\ref{tab:5} report the p-values of the tests and the level of significance of the differences (at a  level of confidence of 95\%). As one can see, the only parameter distributions that are significantly different are the surface gravity, which is slightly lower in our work compared than in Exoplanets.org and SWEET-Cat. The statistical test also confirms that the difference is more significant comparing our data with SWEET-Cat than with Exoplanets.org (p-value ~ 0.0008 instead of 0.0195). Considered as a whole, therefore, these tests suggest that our results are quite comparable with those reported in the literature. 

\begin{table}
 \centering
 \caption[]{\small{Comparison with literature}}
\label{tab:5}
  \vspace{0.25cm} 
 \begin{small}
\begin{tabular}{cccccccccc}
      \toprule      
      & & \multicolumn{2}{c}{TIGRE (45 stars)} &  &\multicolumn{2}{c}{Exoplanets} & &p-value& s.l.\\
      \cmidrule(l){3-4}\cmidrule(l){6-7}
      Parameter & Units & median & mean & & median & mean & &  &  \\
      \midrule
 $T_{eff}$	&(K)					&	$6025$	&	$5975$ 	&&	$6095$	& $5952$	&&$0.7679$	&	ns	 \\
 $\log g$		& 						&	$4.06$		&	$4.18$ 	&&	$4.26$		& $4.28$		&&$0.0195$	&	*	 \\
 $[Fe/H]$	 	& 						&	$0.07$		&	$0.12$ 	&&	$0.20$		& $0.18$  	&&$0.1638$ 	&	ns  \\
 $V \sin i$	&(km/s)			&	$5.55$		&	$7.39$ 	&&	$4.10$		& $6.92$		&&$0.2732$	&	ns		\\	
 $M_{*}$	    &(M$_\odot$)	&	$1.18$		&	$1.13$ 	&&	$1.22$		& $1.19$		&&$0.1010$	&	ns  \\
 $R_{*}$	    &(R$_\odot$)	&	$1.57$		&	$1.47$ 	&&	$1.34$		& $1.36$		&&$0.0868$	&	ns   \\
      \toprule      
      & & \multicolumn{2}{c}{TIGRE (44 stars)} &  &\multicolumn{2}{c}{SWEET-Cat} & &p-value& s.l.\\
      \cmidrule(l){3-4}\cmidrule(l){6-7}
      Parameter & Units & median & mean & & median & mean & &  &  \\
      \midrule
$T_{eff}$		& (K)				&	$6046$	&	$5977$ 	&&	$6133$	& $6036$	&&$0.2936$	&	ns	 \\
 $\log g$		& 						&	$4.06$		&	$4.18$ 	&&	$4.33$		& $4.34$		&&$0.0008$	&	***	 \\
 $[Fe/H]$	 	& 						& $0.08$		& $0.12$		&&	$0.20$		&	$0.17$ 	&&$0.1973$ 	&	ns  \\
 $M_{*}$	    &(M$_\odot$)	&	$1.18$		&	$1.13$ 	&&	$1.24$		& $1.18$		&&$0.0700$	&	ns  \\
 \hline
\end{tabular} 
\end{small}
 \end{table}
 

 \begin{figure}
\includegraphics[width=1\linewidth, angle=0]{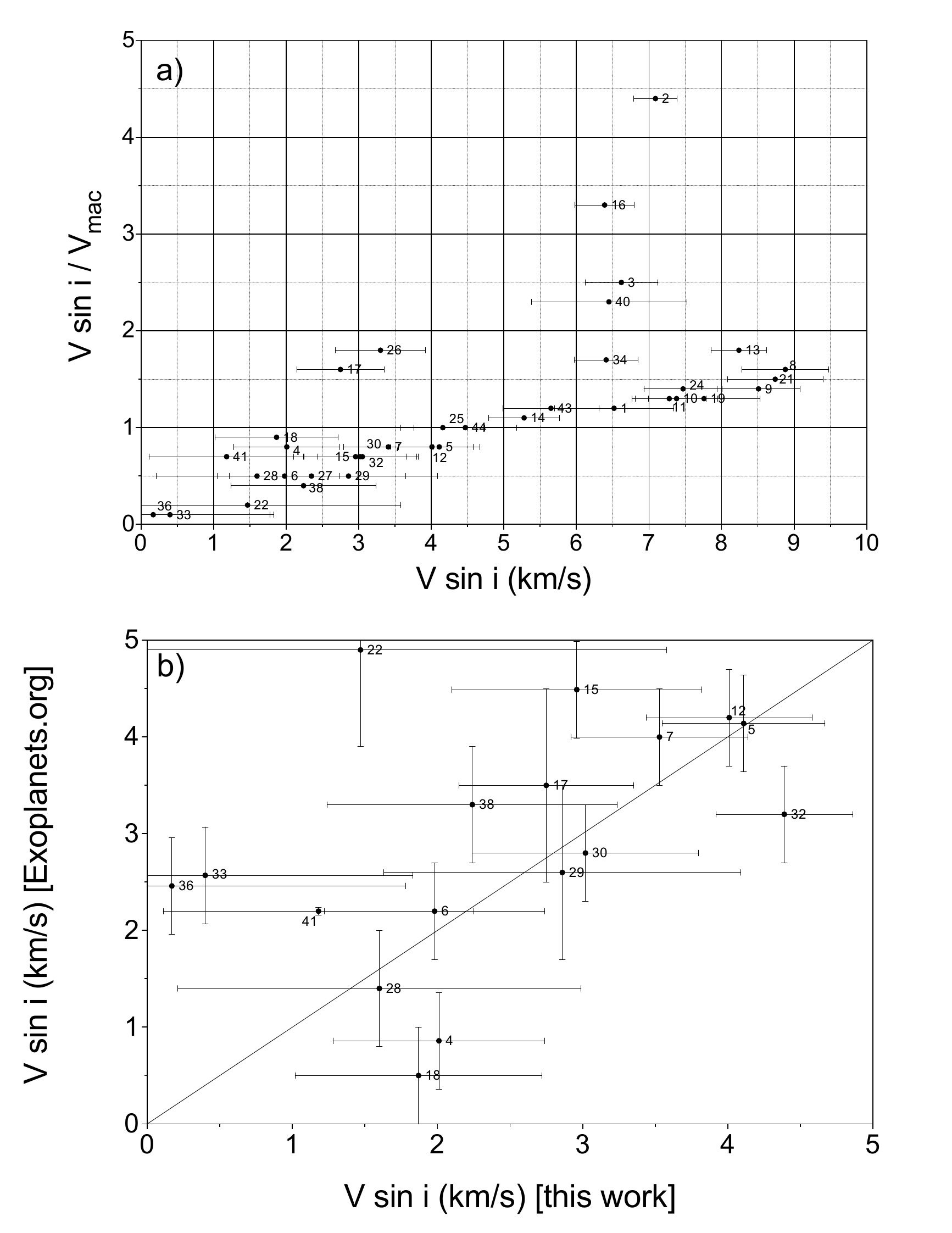}
\caption{\textit{a) The ratio $V \sin i / V_{mac}$ as a function of $V \sin i$. Below  $V \sin i = 4$ the ratios are lower than 1. Three stars, 33, 36 and 22 have \textsf{iSpec} have values with uncertainties that include zero. b) Zoom of the region in Fig.~\ref{Comp}g with $V \sin i < 5$ km/s.}}
\label{VsiniVmac}
\end{figure}

As we mentioned before, as the temperature of the stars goes down, $V_{mic}$ and $V_{mac}$ become comparable to $V \sin i$ and thus it is more complicated to separated one from the others. In \citet{Valenti2005}  spectral synthesis analysis, the authors recognized this problem stating, in particular, that, ``...adopting a global macroturbulence relationship should yield more accurate results than solving for $V_{mac}$ in each individual spectrum.'' To determine such relation they fixed $V \sin i = 0$, obtaining the maximum values $V_{mac}$ could have at different temperatures. Note that these authors did not report any dependence on the spectral resolution, although they used spectra with R between 50,000 and almost 100,000. The maximum relation they deduced can be seen in Fig.~\ref{Vmac}. According to these authors, below $T_{eff} = 5800$ K $V \sin i$ becomes negligible, and what we measure then must be the “real” $V_{mac}$. However, this conclusion contradicts what was expected based on the semi-empirical relation established by \citet{Gray1984b} and later the minimum relation for $V_{mac}$ obtained by \citet{Bruntt2010} by line modeling (the two relations can also be seen in Fig.~\ref{Vmac}). These results suggest that applying the right macro (and micro) turbulence relationship one could obtain a value of $V \sin i \neq 0$ below $T_{eff} = 5800$ K. In fact, in our analysis of the Sun, we did reproduced the value of $V \sin i$, using the relation for $V_{mic}$ obtained by \citet{Tsantaki2014} and $V_{mac}$  determined by \citet{Doyle2014}, both depending not only on $T_{eff}$ but also on on $\log g$, and where $V \sin i < V_{mac}$. The question then is how low can $V \sin i$ be compared to $V_{mac}$ and still be distinguishable by \textsf{iSpec}? 

In Fig.~\ref{VsiniVmac}a, we compare $V \sin i$  with the ratio $V \sin i / V_{mac}$. What we observe is that below $V \sin i = 4$ km/s the ratio is lower than one. As one can see in Fig.\ref{Vmac}, a value of  $V_{mac} = 4$ km/s ($V \sin i / V_{mac} = 1$) corresponds to $T_{eff} \sim 5800$ K. Therefore, our results are at the same time consistent with the conclusion of \citet{Valenti2005}, since below $T_{eff} = 5800$ K $V \sin i$ is lower than $V_{mac}$, and consistent with \citet{Gray1984a}, \citet{Bruntt2010} and \citet{Doyle2014}, since $V \sin i \ne 0$. But how low could a value of $V \sin i$ below $V_{mac}$ be? This question we already answered in Fig.~\ref{Comp}g where we compared our values of $V \sin i$ with the values reported by Exoplanets.org. To get a better view, in Fig.~\ref{VsiniVmac}b we zoom in on values of $V \sin i \le 5$ km/s. Except for three stars, 22, 33 and 36, with $V \sin i / V_{mac} < 0.4$, all the other stars have $V \sin i$ comparable to the values reported in Exoplanets.org (in fact, two of the stars, 4 and 16, have higher values). In Fig.~\ref{VsiniVmac}a note that the \textsf{iSpec} uncertainty increases as $V \sin i$ goes down. As a consequence, the possible values for the stars 22, 33 and 36 include zero. However, could stars with $V \sin i = 0$ exist physically? Considering that the loss of angular momentum plays an important role in the formation of stars this would seem difficult to explain  (note that we did obtained $V \sin i = 0$ for some stars in our initial list, but they were not included in our study). Since \citet{Gray1984b} study the problem seems clear: how can we measure the rotation of a star where $V_{mac}$ is as high or even higher than $V \sin i$? It seems that the best approach is assuming an a priori global relation and see what comes out from the residual  \citep{Gray1984a,Gray1984b,Fischer2005b,Bruntt2010,Tsantaki2014,Doyle2014}. However, to stay safe, due to their higher uncertainties we should not consider 22, 33 and 36 in our statistical analysis for $V \sin i$.  


\begin{figure}
\includegraphics[width=0.7\linewidth, angle=270]{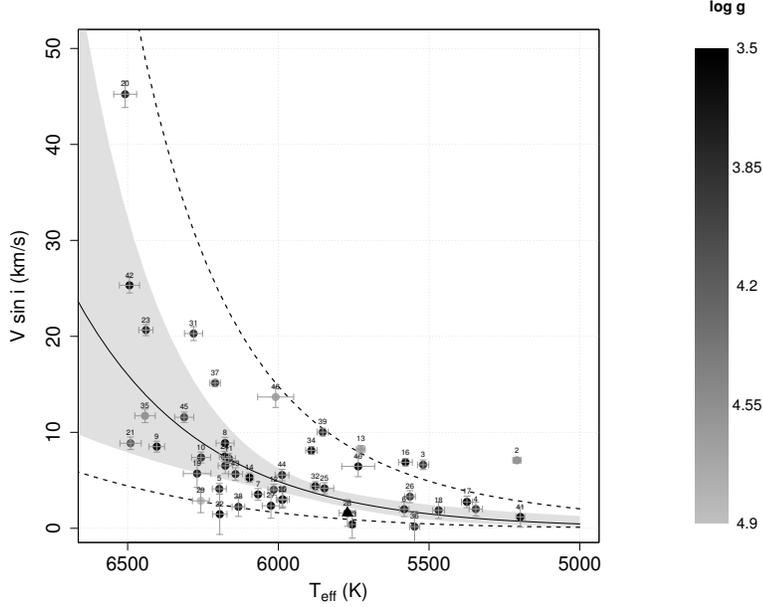}
\caption{\textit{Exponential relation between the rotational velocity, the temperature and the surface gravity for the 46 stars in our sample. The black triangle represent the Sun. The gray area corresponds to the interval of confidence and the dashed curves delimit the prediction interval.}}
\label{Vsini}
\end{figure}

Considering the results above, one expects the rotational rotation to decrease with the temperature, but still be above zero in cool stars. Moreover, since \citet{Tsantaki2014} and \citet{Doyle2014} have found relations for $V_{mic}$ and $V_{mac}$  that depend not only on $T_{eff}$ but also on on $\log g$, we might expect a similar relation for $V \sin i$, $T_{eff}$ and $\log g$. In Fig.~\ref{Vsini}, we show the diagram of $V \sin i$ and  $T_{eff}$ for our stars. The dependence in $\log g$ is shown by the gray-scale bar. In Fig.~\ref{Vsini}, we also traced over our data the bi-exponential relation we obtained, together with the interval of confidence (in gray) and  the prediction interval (dashed curves), which takes into account the uncertainty of each measurement. The final relation we obtained is the following:  
\begin{equation}\label{eq8}
\frac{V \sin i}{\rm km/s} = \exp \left[\ A \left( \frac{T_{eff}}{1000{\rm K}} \right)  + B \log g - C\  \right]
\end{equation}
where $A = 2.20 \pm 0.36$, $B = 0.30 \pm 0.46$ and $C = 12.91 \pm 3.59$, 
and which has a multiple correlation coefficient of $r^2 = 0.6329$. Except for the stars 2, 3, 13 and 16, and the three stars  with highest uncertainties (22, 33, and 36; not considered in this relation) all our data fit well inside the prediction interval. 

\begin{figure}
\includegraphics[width=0.7\linewidth, angle=270]{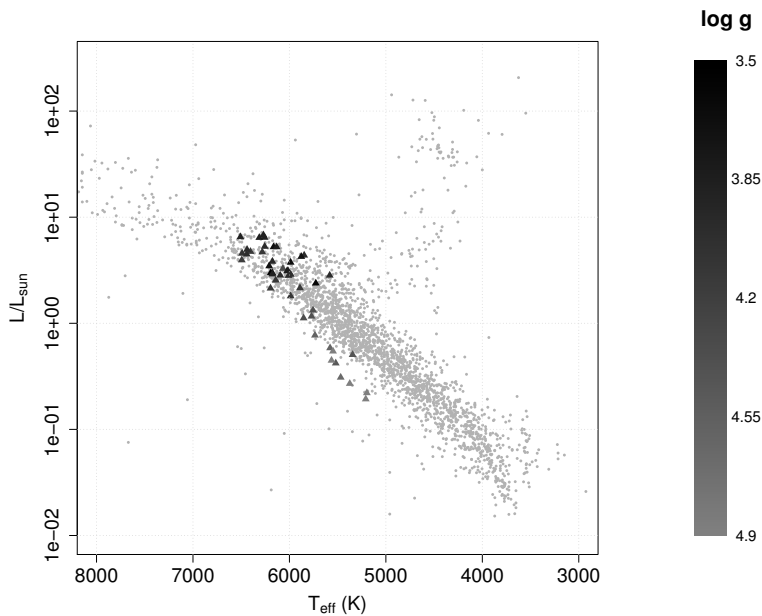}
\caption{\textit{HR diagram of our 46 stars, overlayed on the main sequence of Hipparchos stars.}}
\label{HR}
\end{figure}

Note that in order to obtain the highest correlation coefficient possible, 8 stars suspected to have peculiarly high rotation for their temperature were considered as outliers. They are, from right to left in Fig.~\ref{Vsini}: 2, 3, 16, 13, 40, 39, 46 and 37. Different reasons were explored that could explain why these stars would be outliers. One is the age of the stars \citep[e.g., ][]{Stauffer1986}, younger stars rotating faster than older stars \citep[see Fig.~1.6 in][]{Tassoul2000}. In \citet[][]{Tassoul2000} it was also shown that young stars trace the same relation of $V \sin i$ with $T_{eff}$ as old stars, with only higher velocities, forming an upper sequence (or upper envelope). 
This could be what we see in Fig.~\ref{Vsini}. However, in Fig.~\ref{HR} the HR diagram for our stars compared to Hipparchos stars suggests that, except for three stars with slightly higher luminosity for their temperature (6, 25 and 32; none of these stars forming the envelope) all of the stars more luminous than the Sun are clearly on the main sequence. This eliminates the young age hypothesis. Another explanation could be peculiar surface activity.  Since more than one phenomena can cause such activities, the expected effect would be pure random dispersion. Revising the literature for each of the star in our sample we did find 8 stars with reported peculiarities: 2, 3, 17, 26, 33, 37, 39 and 46. The type of peculiarities encountered included, ``Flare star'', ``Rotationally variable'', ``Variable BY DRa'', and ``Double or Multiple star''. Of these``active'' stars only five in Fig.~\ref{Vsini} have an higher $V \sin i$ for their temperature:  2, 3, 37, 39 and 46. This left three stars (13, 16  and 40) with unexplained, relatively high $V \sin i$ values. In fact, checking their $V_{mac}$ we found these stars have lower values than stars with comparable temperatures. However, in our various attempts to get the higher highest correlation coefficient possible, we judged better to keep them as outliers.

\begin{figure}[h!]
\includegraphics[width=0.7\linewidth, angle=270]{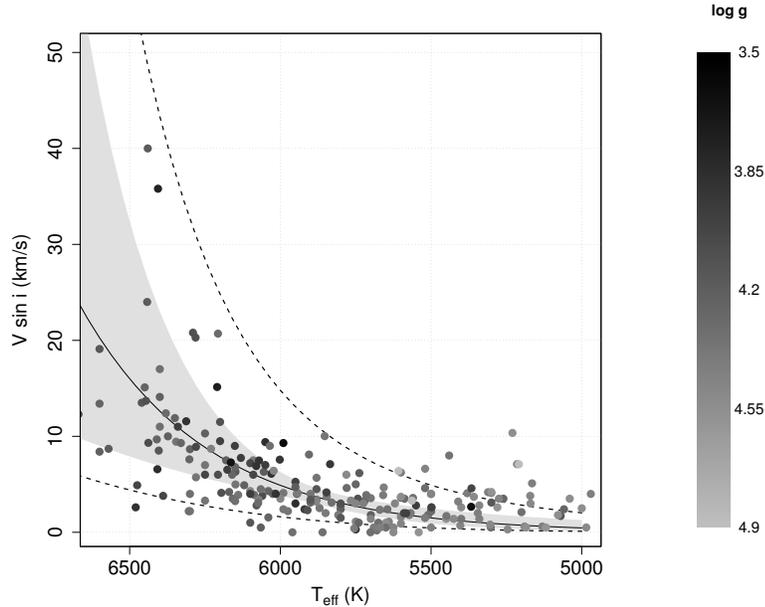}
\caption{\textit{Data as found in Exoplanets.org.}}
\label{f3}
\end{figure}

To verify our relation, in Fig.~\ref{f3} we traced it over the distribution of the rotational velocities and temperatures of the stars that were in our initial sample based on Exoplanets.org. As one can see, except for a few stars below $T_{eff} = 5500$ K, with higher velocities, and stars below $T_{eff} = 5500$ K, with lower velocities (some with $V \sin i = 0$) the majority of the stars in this sample fall well between the prediction interval of our empirical relation. This result suggests that the decrease in angular momentum of low mass stars is a non-aleatory phenomenon, most probably reflecting the action of one specific mechanism, like, for example, magnetic braking or stellar wind \citep{Wolff1997, Tassoul2000, Uzdensky2002}. An exciting possibility, however, could be that this relation somehow is coupled to the formation of the planets. Although this hypothesis proposed in the late 1960s was rapidly rejected, since no planet out of the solar system was known at the time, the discovery of exoplanets allows us today to test this idea anew \citep[e.g.,][]{Berget2010}. This will be the subject of Paper~{\rm II}, in search of a connection between the formation of stars and planets. 


 
\section{Conclusions}
\label{conclusion}

In this study we have shown that our method of analysis developed for the TIGRE telescope using iSpec on intermediate Echelle resolution spectra yields results about the physical characteristics of stars, host of exoplanets, that are comparable to those obtained using bigger telescopes and standard spectra analysis methods, using high resolution spectra. Our results show that TIGRE can have an helpful contribution in the follow up of exoplanet surveys around bright stars, like TESS and PLATO. Such follow up studies are essential in order to understand how the formation of planets is connected to the formation of their host stars \citep{Eisner2020}. 
 
\acknowledgments
We like to thank an anonymous referee for his careful revision of our results and for his comments and suggestions that helped us improved our work. L. M. F. T. would like to thank the CONACyT for its support through the grant CVU 555458. She also acknowledges CONACyT for travel support (bilateral Conacyt-DFG projects 192334, 207772, and 278156), as well as the Universidad de Guanajuato (Direcci\'on de Apoyo a la Investigaci\'on y al Posgrado, DAIP, and Campus Guanajuato) for support given for conference participation and international collaborations. L. M. F. T. also thanks the time request committee of the TIGRE for granting her the observations and the whole observing team for their support in getting the data that were used in this study. More personally, she thanks Sebastian Kohl for his help with \textsf{MOLECFIT}. This research has made use of the Exoplanet Orbit Database, the Exoplanet Data Explorer at exoplanets.org \citep{Han2014}, the exoplanets.eu \citep{Schneider2011} and the NASA's Astrophysics Data System. 

\newpage
\begin{appendices}
   
\section{List of spectral lines and segments defined for our analysis}
    \label{sec:ap-A}
{
\scriptsize
    \begin{center}
\begin{longtable}{lccccc}
\caption[Lines and segments defined in this work]{Lines and segments defined in this work.} \label{tab:B1} \\

\hline \multicolumn{1}{|c|}{\textbf{Line}} & \multicolumn{1}{c|}{\textbf{Wave Peak}} & \multicolumn{1}{c|}{\textbf{Wave Base}} & \multicolumn{1}{c|}{\textbf{Wave Top}} & \multicolumn{1}{c|}{\textbf{Segm. Wave base}} & \multicolumn{1}{c|}{\textbf{Segm. Wave top}}  \\ \hline 
\endfirsthead

\multicolumn{6}{c}%
{{\bfseries \tablename\ \thetable{} -- continued from previous page}} \\
\hline \multicolumn{1}{|c|}{\textbf{Line}} & \multicolumn{1}{c|}{\textbf{Wave Peak}} & \multicolumn{1}{c|}{\textbf{Wave Base}} & 
\multicolumn{1}{c|}{\textbf{Wave Top}} & 
\multicolumn{1}{c|}{\textbf{Segm. Wave base}} & \multicolumn{1}{c|}{\textbf{Segm. Wave top}}  \\ \hline 
\endhead

\hline \multicolumn{6}{|c|}{{Continued on next page}} \\ \hline
\endfoot

\hline \hline
\endlastfoot

Na 1 & 588.9959 & 588.9422 & 589.0422 & 588.8922 & 589.0922 \\
Na 1 & 589.5916 & 589.5411 & 589.6411 & 589.4911 & 589.6911 \\
Fe 1 & 593.0186 & 592.9859 & 593.0539 & 592.9359 & 593.1039 \\
Fe 1 & 593.4665 & 593.4289 & 593.5059 & 593.3789 & 593.5559 \\
No ident. & 595.6706 & 595.6206 & 595.7206 & 595.5706 & 595.7706 \\
Fe 1 & 597.5341 & 597.4898 & 597.5898 & 597.4398 & 597.7729 \\
Fe 1 & 597.6777 & 597.6299 & 597.7229 & - & - \\
Fe 1 & 598.4831 & 598.4319 & 598.5689 & 598.3819 & 598.8099 \\
Fe 1 & 598.7088 & 598.6449 & 598.7599 & - & - \\
Fe 1 & 600.2986 & 600.2519 & 600.3509 & 600.2019 & 600.4009 \\
Fe 1 & 600.8552 & 600.8249 & 600.8989 & 600.7749 & 600.9489 \\
Mn 1 & 601.6628 & 601.6110 & 601.7110 & 601.5610 & 601.7610 \\
Fe 1 & 602.0142 & 601.9637 & 602.0637 & 601.9137 & 602.1137 \\
Fe 1 & 602.4066 & 602.3579 & 602.4639 & 602.3079 & 602.5139 \\
Fe 1 & 605.6032 & 605.5599 & 605.6809 & 605.5099 & 605.7309 \\
Fe 1 & 606.5494 & 606.5009 & 606.5919 & 606.4509 & 606.6419 \\
Fe 1 & 607.8490 & 607.7729 & 607.8769 & 607.7229 & 607.9269 \\
Fe 1 & 608.2757 & 608.2180 & 608.3180 & 608.1680 & 608.3680 \\
Fe 1 & 608.5228 & 608.4775 & 608.5775 & 608.4275 & 608.6275 \\
Ca 1 & 612.2225 & 612.1703 & 612.2703 & 612.1203 & 612.3203 \\
No ident. & 615.1608 & 615.1108 & 615.2108 & 615.0608 & 615.2608 \\
Ca 1 & 616.2171 & 616.1690 & 616.2690 & 616.1190 & 616.3190 \\
Fe 1 & 617.0503 & 617.0028 & 617.1168 & 616.9528 & 617.1668 \\
Fe 1 & 617.3340 & 617.2828 & 617.3838 & 617.2328 & 617.4338 \\
Fe 1 & 621.3421 & 621.2988 & 621.3958 & 621.2488 & 621.4458 \\
Fe 1 & 621.9270 & 621.8418 & 621.9698 & 621.7918 & 622.0198 \\
Fe 1 & 623.0722 & 623.0278 & 623.1868 & 622.9778 & 623.3578 \\
Fe 1 & 623.2644 & 623.1868 & 623.3078 & - & - \\
Fe 1 & 624.6326 & 624.5898 & 624.6868 & 624.5398 & 624.7368 \\
Fe 1 & 625.2564 & 625.2108 & 625.3108 & 625.1608 & 625.7298 \\
Fe 1 & 625.4240 & 625.3298 & 625.5098 & - & - \\
Fe 1 & 625.6343 & 625.5628 & 625.6798 & - & - \\
Fe 1 & 629.0951 & 629.0473 & 629.1473 & 628.9973 & 629.1973 \\
Fe 1 & 629.7808 & 629.7138 & 629.8548 & 629.6638 & 629.9048 \\
Fe 1 & 630.1508 & 630.0898 & 630.2028 & 630.0398 & 630.3528 \\
Fe 1 & 630.2514 & 630.2028 & 630.3028 & - & - \\
Fe 1 & 632.2710 & 632.2228 & 632.3128 & 632.1728 & 632.3628 \\
Fe 1 & 633.5331 & 633.4658 & 633.5888 & 633.4158 & 633.7978 \\
Fe 1 & 633.6827 & 633.6388 & 633.7478 & - & - \\
Fe 1 & 635.5038 & 635.4468 & 635.5768 & 635.3968 & 635.6268 \\
Fe 1 & 635.8671 & 635.8128 & 635.9258 & 635.7628 & 635.9758 \\
Fe 1 & 638.0743 & 638.0264 & 638.1264 & 637.9764 & 638.1764 \\
Fe 1 & 639.3612 & 639.2968 & 639.4278 & 639.2468 & 639.4778 \\
Fe 1 & 640.8011 & 640.7578 & 640.9138 & 640.7078 & 640.9638 \\
Fe 1 & 641.1646 & 641.0878 & 641.2198 & 641.0378 & 641.2698 \\
Fe 2 & 641.6962 & 641.6449 & 641.7449 & 641.5949 & 641.7949 \\
Fe 1 & 641.9949 & 641.9428 & 642.0408 & 641.8928 & 642.2598 \\
Fe 1 & 642.1377 & 642.0758 & 642.2098 & - & - \\
Fe 1 & 643.0851 & 643.0158 & 643.1528 & 642.9658 & 643.3681 \\
Fe 2 & 643.2663 & 643.2181 & 643.3181 & - & - \\
Ca 1 & 643.9063 & 643.8572 & 643.9572 & 643.8072 & 644.0072 \\
Fe 2 & 645.6405 & 645.5866 & 645.6866 & 645.5366 & 645.7366 \\
Ca 1 & 646.2606 & 646.2081 & 646.3081 & 646.1581 & 646.3581 \\
Fe 1 & 646.9200 & 646.8711 & 646.9711 & 646.8211 & 647.0211 \\
Fe 1 & 647.5657 & 647.5117 & 647.6117 & 647.4617 & 647.6617 \\
Fe 1 & 648.1882 & 648.1362 & 648.2362 & 648.0862 & 648.2862 \\
Fe 1 & 649.4989 & 649.4197 & 649.5437 & 649.3697 & 649.5937 \\
Fe 2 & 651.6098 & 651.5587 & 651.6587 & 651.5087 & 651.7087 \\
Fe 1 & 651.8385 & 651.7868 & 651.8868 & 651.7368 & 651.9368 \\
Fe 1 & 654.6245 & 654.5757 & 654.6967 & 654.5257 & 654.7467 \\
H 1 & 656.2808 & 655.5483 & 656.6832 & 655.1934 & 656.7340 \\
Fe 1 & 657.5003 & 657.4507 & 657.5507 & 657.4007 & 657.6007 \\
Fe 1 & 659.3887 & 659.3417 & 659.4537 & 659.2917 & 659.5037 \\
Fe 1 & 659.7585 & 659.7073 & 659.8073 & 659.6573 & 659.8573 \\
Fe 1 & 660.9067 & 660.8605 & 660.9605 & 660.8105 & 661.0105 \\
Ni 1 & 664.3626 & 664.3139 & 664.4139 & 664.2639 & 664.4639 \\
Fe 1 & 667.7983 & 667.7297 & 667.8707 & 667.6797 & 667.9207 \\
Fe 1 & 670.5134 & 670.4570 & 670.5570 & 670.4070 & 670.6070 \\
No ident. & 671.3073 & 671.2573 & 671.3573 & 671.2073 & 671.4073 \\
Ca 1 & 671.7701 & 671.7138 & 671.8138 & 671.6638 & 671.8638 \\
Fe 1 & 672.6657 & 672.6178 & 672.7178 & 672.5678 & 672.7678 \\
Fe 1 & 675.0182 & 674.9653 & 675.0653 & 674.9153 & 675.1153 \\
Fe 1 & 680.6856 & 680.6358 & 680.7358 & 680.5858 & 680.7858 \\
Fe 1 & 682.0359 & 681.9894 & 682.0894 & 681.9394 & 682.1394 \\
Fe 1 & 682.8620 & 682.8085 & 682.9085 & 682.7585 & 682.9585 \\
No ident. & 683.9811 & 683.9311 & 684.0311 & 683.8811 & 684.0811 \\
Fe 1 & 684.3658 & 684.3150 & 684.4150 & 684.2650 & 684.4650 \\
Fe 1 & 691.6669 & 691.6218 & 691.7218 & 691.5718 & 691.7718 \\
No ident. & 693.3635 & 693.3135 & 693.4135 & 693.2635 & 693.4635 \\
Fe 1 & 694.5196 & 694.4703 & 694.5703 & 694.4203 & 694.6203 \\
- & - & - & - & 694.6410 & 694.8410 \\
Fe 1 & 695.1251 & 695.0721 & 695.1721 & 695.0221 & 695.2221 \\
Fe 1 & 703.8209 & 703.7718 & 703.8718 & 703.7218 & 703.9218 \\
Fe 1 & 706.8440 & 706.7918 & 706.8918 & 706.7418 & 706.9418 \\
Fe 1 & 709.0378 & 708.9850 & 709.0850 & 708.9350 & 709.1350 \\
Fe 1 & 713.0900 & 713.0451 & 713.1451 & 712.9951 & 713.1951 \\
Fe 1 & 713.3001 & 713.2453 & 713.3453 & 713.1953 & 713.3953 \\
CN 1 & 714.5241 & 714.4768 & 714.5768 & 714.4268 & 714.6268 \\
Ca 1 & 714.8155 & 714.7666 & 714.8666 & 714.7166 & 714.9166 \\
Fe 1 & 715.5670 & 715.5125 & 715.6125 & 715.4625 & 715.6625 \\
No ident. & 716.4473 & 716.3185 & 716.5085 & 716.2685 & 716.5585 \\
Fe 1 & 717.5970 & 717.5403 & 717.6403 & 717.4903 & 717.6903 \\
Fe 1 & 721.9712 & 721.9134 & 722.0134 & 721.8634 & 722.2190 \\
CN 1 & 722.1100 & 722.0690 & 722.1690 & - & - \\
No ident. & 724.4812 & 724.4312 & 724.5312 & 724.3812 & 724.5812 \\
Fe 1 & 732.0693 & 732.0178 & 732.1178 & 731.9678 & 732.1678 \\
Fe 1 & 738.6353 & 738.5818 & 738.6818 & 738.5318 & 738.7318 \\
Fe 1 & 738.9363 & 738.8454 & 738.9974 & 738.7954 & 739.0474 \\
Fe 1 & 741.1151 & 741.0394 & 741.1764 & 740.9894 & 741.2264 \\
Ni 1 & 742.2264 & 742.1770 & 742.2770 & 742.1270 & 742.3270 \\
No ident. & 744.0877 & 744.0377 & 744.1377 & 743.9877 & 744.1877 \\
Fe 1 & 744.5740 & 744.5174 & 744.6654 & 744.4674 & 744.7154 \\
Fe 1 & 749.5088 & 749.4484 & 749.5724 & 749.3984 & 749.6224 \\
Fe 1 & 751.1024 & 751.0024 & 751.1854 & 750.9524 & 751.2354 \\
Fe 1 & 771.0389 & 770.9827 & 771.0827 & 770.9327 & 771.1327 \\
Fe 1 & 772.3237 & 772.2724 & 772.3724 & 772.2224 & 772.4224 \\
Fe 1 & 774.8304 & 774.7653 & 774.8613 & 774.7153 & 774.9113 \\
Ni 1 & 775.1163 & 775.0625 & 775.1625 & 775.0125 & 775.2125 \\
Fe 1 & 778.0562 & 777.9613 & 778.1263 & 777.9113 & 778.1763 \\
Fe 1 & 783.2221 & 783.1453 & 783.3183 & 783.0953 & 783.3683 \\
Fe 1 & 793.7145 & 793.6112 & 793.7802 & 793.5612 & 793.8302 \\
Fe 1 & 794.5839 & 794.5132 & 794.6502 & 794.4632 & 794.7002 \\
Fe 1 & 799.8967 & 799.8112 & 799.9622 & 799.7612 & 800.0122 \\
No ident. & 804.6052 & 804.5282 & 804.7002 & 804.4782 & 804.7502 \\
No ident. & 808.5170 & 808.4442 & 808.6012 & 808.3942 & 808.6512 \\
Fe 1 & 820.7791 & 820.7284 & 820.8284 & 820.6784 & 820.8784 \\
Fe 1 & 832.7062 & 832.6341 & 832.7711 & 832.5841 & 832.8211 \\
Fe 1 & 838.7760 & 838.7061 & 838.8521 & 838.6561 & 838.9021 \\
Fe 1 & 846.8392 & 846.7820 & 846.8930 & 846.7320 & 846.9430 \\
Fe 1 & 851.4073 & 851.3290 & 851.4650 & 851.2790 & 851.5150 \\
Fe 1 & 868.8639 & 868.7760 & 868.9430 & 868.7260 & 868.9930 \\
No ident. & 871.0395 & 870.9895 & 871.0895 & 870.9395 & 871.1395 \\

\hline
\end{longtable}
\end{center}
}

  \end{appendices}
  

\end{document}